\documentclass[preprint,a4paper,fleqn,finallayout]{w-art}
\makeatletter
 \def\@oddfoot{}%
  \def\@evenfoot{}%
  \@abstractcopyrightfalse
\makeatother
\usepackage{times,cite,w-thm}
\usepackage{braket}
\usepackage{bbm}
\usepackage{color}
\usepackage{alltt}
\newcommand{\ketbra}[2]{\ket{#1}\bra{#2}}
\renewcommand{\braket}[2]{\bra{#1}\left.#2\right>}

\theoremstyle{plain}

\theoremstyle{definition}

\usepackage[]{graphicx}

\newcommand{\ie}{i.e.}
\newcommand{\eg}{e.g.}
\newcommand{\plotwidth}{0.75\textwidth}

\newcommand{\updown}{\updownarrow}
\newcommand{\leftright}{\leftrightarrow}
\newcommand{\shor}{\ket{\updown}}
\newcommand{\svert}{\ket{\leftright}}
\newcommand{\sdiaga}{\frac{1}{\sqrt{2}}(\shor + \svert)}
\newcommand{\sdiagb}{\frac{1}{\sqrt{2}}(\shor - \svert)}

\begin{document}
\DOIsuffix{theDOIsuffix}
\Volume{16}
\Month{01}
\Year{2007}
\pagespan{1}{}
\Receiveddate{30 November 2007}

\title[Quantum key distribution]{Recent developments in quantum key
distribution:\\ theory and practice}

\author[W.~Mauerer]{Wolfgang Mauerer\footnote{Corresponding
    author\quad E-mail:~\textsf{wolfgang.mauerer@ioip.mpg.de}}}
\address{Max Planck Research Group, Institute of Optics, Information
  and Photonics, Integrated Quantum Optics Group, University of
  Erlangen-N{\"u}rnberg, G{\"u}nther-Scharowsky-Stra{\ss}e 1/Bau 24, 91058
  Erlangen, Germany}
\author[W.~Helwig]{Wolfram Helwig}
\author[Ch.~Silberhorn]{Christine Silberhorn}

\dedicatory{In honour of Max Planck (1858--1947) on the occasion of his 150th
  birthday.}

\begin{abstract}
  Quantum key distribution is among the foremost applications of quantum
  mechanics, both in terms of fundamental physics and as a technology on the
  brink of commercial deployment.  Starting from principal schemes and initial
  proofs of unconditional security for perfect systems, much effort has gone
  into providing secure schemes which can cope with numerous experimental
  imperfections unavoidable in real world implementations. In this paper, we
  provide a comparison of various schemes and protocols. We analyse their
  efficiency and performance when implemented with imperfect physical
  components. We consider how experimental faults are accounted for using
  effective parameters. We compare various recent protocols and provide
  guidelines as to which components propose best advances when being improved.
\end{abstract}

\maketitle
\renewcommand{\leftmark}{W.~Mauerer et al.: Quantum key distribution}
\renewcommand{\rightmark}{\leftmark}

\section{Introduction}
During the last decade, quantum key distribution (QKD) was established among
the most active and productive research fields in quantum physics. Sharing
secret information between remote parties is not only a practical problem, but
has interesting outreach into numerous fields ranging from information theory
to fundamental physics. Although the fundamental ideas of QKD were already
laid out around 1970 by Stephen Wiesner, they were not published until a
decade later, in 1983~\cite{Wiesner1983}. The year 1984 brought the seminal
proposal of Charles Bennett and Gilles Brassard in form of the BB84 key
distribution protocol~\cite{Bennett1984}. It was followed by numerous
alternative variants and improvements
(\eg,~\cite{Bennett1992,Ekert1991,Scarani2004,Hwang2003}), but the methods
usually remain closely related to the original proposal.

The basic idea of QKD is as follows. Two parties, which we name Planck and
Bohr (it is also common to refer to them as Alice and Bob), establish a shared
secret key between them because they want to discuss sensible facts about
their theories. The malicious eavesdropper Einstein (alternatively called Eve
in a cryptographic context) who possesses unlimited technological power --
only restricted by the laws of quantum mechanics -- does not believe in
Planck's and Bohr's theory. He would like to listen to their communication in
order to prove their arguments wrong.  Fortunately, their quantum theory is
right, and he is not able to gain any information about the shared secret
key. This key is used to encode information, and the encoding can be shown to
be perfectly secure provided the key is only known to Planck and Bohr. Thus
quantum cryptography is referred to as unconditionally secure.

Security proofs for QKD schemes are solely based on the laws of
physics and do not use any assumptions about computational complexity
-- in contrast to purely classical schemes. Real world effects like
lossy transmission or imperfect detectors are unavoidable in
practice. Nevertheless, security can also be proved under these
circumstances. The obtainable key rates and achievable transmission
distance highly depend on the utilised hardware. It is important to
select components such that optimal performance is achieved. In this
paper, we will present a survey of a number of available protocols. We
will also discuss how imperfections affect their performance, and how
to choose the best hardware. This includes advice on which components
need to be optimised in various practical settings to obtain best
results.

\begin{figure}
  \centering\includegraphics[width=0.75\textwidth]{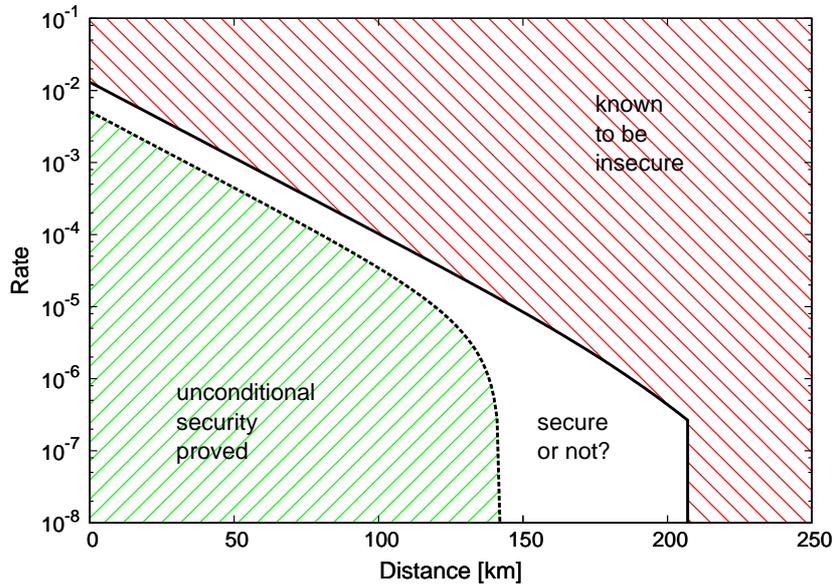}
  \caption{Typical performance diagram for a QKD system. The
    dashed line represents a lower bound on the secret key rate. This means it
    can be proven that at a given distance, \emph{at least} this number of
    secret bits can be transmitted per pulse. The solid line provides
    fundamental physical upper bounds, and it is known that no secret key can
    be established beyond this distance respectively that no larger number of
    secret bits per transmitted pulse can be obtained. The region between the
    lower and upper bound is unknown terrain: It is still undecided if secure
    QKD is possible in this area. The aim of current research is therefore to
    increase the lower bounds
    (\eg,~\cite{Renner2005,Kraus2005,Fung2006,Hwang2003}) and decrease the
    upper bounds (\eg,~\cite{Moroder2006,Moroder2006a,Kraus2005}).
    Section~\ref{sec:tmd_decoy} introduces one scheme which nearly closes this
    gap. The key rate is given in secret bits per pulse, and must be
    multiplied with the repetition rate of the source to obtain the proper
    data rate. With current technology, the maximal repetition rate is not
    limited by the source, but mainly by the dead time of detectors,
    cf.~Ref.~\cite{Coldenstrodt-Ronge2007}.}\label{fig:dist_example}
\end{figure}

General reviews about QKD can be found in
References~\cite{Duvsek2006,Gisin2002}, and we will therefore not try
to provide a detailled elementary introduction in this paper, but
refer the reader to the mentioned references to become qcquainted with
the basic concepts. The principal setting of every QKD protocol
between two parties is depicted in Fig.~\ref{fig:qkd_overview}. The
performance of QKD systems is usually visualised in terms of diagrams
which plot the secret key rate over transmission distance.
Fig.~\ref{fig:dist_example} shows an example, and the reader is
referred to the caption for information on how to interpret such a
diagram.

\section{QKD hardware}
We now turn our attention to the hardware used for signal generation,
transmission and reception. Security proofs require idealised components.
Unavoidable experimental imperfections are handled by introducing effective
error parameters, and we will concentrate on these in the
following.\footnote{Obviously imperfect components must be described by a
  physical model. Aditional attacks can become feasible -- possibly even
  classical attacks -- if side effects arise which are not included in the
  model.}
Note that Table~\ref{tab:qkd_params} provides
a hardware comparison for recently performed QKD experiments.

Before we commence further, we would like to point to ``equation''~(77)
from Ref.~\cite{Gisin2002}: 
\begin{equation}
\text{Infinite security} \Rightarrow \text{Infinite cost}
\Rightarrow \text{Zero practical interest}.
\end{equation}
This means that practical usability is required for every QKD system --
cryptography is no good if everyone sends everything unencrypted because no
one can afford to use encryption. Therefore, it is especially important
to consider real-world applicability. This means especially cost and
technological feasibility. While there may often be better (research)
alternatives for a given component, we concentrate
on the solutions which are not only available in quantities, but also come at
a reasonable price.

\subsection{Photon sources}
Transmitting quantum signals requires a signal source.  With few
exceptions, the protocols considered in this paper deally require
single photons. This can be achieved in practice, but only at a great
expense and experimental effort, see, for instance,
Ref.~\cite{Keller2004}.  In current practice, two sources for QKD are
conventionally employed: attenuated lasers and parametric
downconversion (PDC) sources.

\subsubsection{Attenuated laser beams}
The quantum state emitted by a laser is a coherent state which can be uniquely
characterised by a complex value \(\alpha\). A representation in the Fock
basis is given by 
\begin{equation}
  \ket{\psi} =
  e^{-\frac{|\alpha|^{2}}{2}}\sum_{n=0}^{\infty}\frac{\alpha^{n}}{\sqrt{n!}}\ket{n}.
\end{equation}
Under the assumption that no phase reference for the state exists once it has
left Planck's control~\cite{Gisin2002}, we can write \(\alpha = \mu
e^{i\theta}\) for \(\mu, \theta \in \mathbbm{R}\), and the corresponding phase
averaged density operator becomes 
\begin{equation}
  \hat{\varrho} =
  \frac{1}{2\pi}\int\text{d}\theta\ket{\mu e^{i\theta}}\bra{\mu e^{i\theta}} =
  \sum_{n}p(n)\ketbra{n}{n}.
\end{equation}
The probability distribution \(p(n)\) is given by
\begin{equation} 
  p(n) =
  e^{-|\alpha|^{2}}\frac{|\alpha|^{2n}}{n!},
\end{equation}
\ie, a Poissonian distribution, and the density operator represents a
classical mixture of photon number states.

To approximate the desired single-photon source, the intensity (\ie, \(\mu\))
is chosen such that the probability of emitting a two-photon Fock state is
low. Unfortunately, with less intensity, vacuum contributions (\ie, Fock
states with zero photons) become more and more frequent. In a laser pulse
based system, it is therefore essential to optimise the intensity such that
key rate and transmission distance are maximised.

\subsubsection{Parametric downconversion}
In a PDC process~\cite{Mandel1995}, an optical medium with nonlinear
susceptibility \(\chi^{(2)}\) is pumped by a laser to generate
photon-number entangled states of the form \(\ket{\psi} \propto
\sum_{n=0}^{\infty}c_{n}\ket{n}_{\text{S}}\ket{n}_{\text{I}}\).  The
subscripts \(\text{S}\) and \(\text{I}\) refer to two spatial modes
which are conventionally denoted by ``signal'' and ``idler''. If the
spectrum of the state is approximately a single frequency mode, the
distribution \(p(n)\) is thermal, \ie, \(c_{n} =
\lambda^{n}\). Current results however indicate that the photon number
distribution will shift to a Poissonian distribution the more
frequencies contribute to the state,
cf.~Ref.~\cite{Avenhaus2007,Tapster1998}. Irregardless of the
statistical distribution, PDC sources provide two distinct advantages
in comparison to lasers.

\begin{itemize}
\item The second spatial mode can be used as a trigger, resulting in a
  heralded photon source~\cite{Alibart2005,U'Ren2005,U'Ren2004}. This enables
  to filter out vacuum contributions, resulting in considerably increased
  maximal secure distance~\cite{Horikiri2005,Horikiri2006,Adachi2007}.
\item In combination with the decoy method (refer to
  Section~\ref{sec:decoy}), several techniques have been devised to gain
  higher transmission rates and distances~\cite{Adachi2007,Mauerer2007}.
\item Entanglement-based protocols as introduced by Ekert in
  1991~\cite{Ekert1991} become possible.
\end{itemize}

Note that the advantage of parametric downconversion comes only into
play if a satisfactory source efficiency can be achieved. This has not
been the case for crystal-based sources which used to be state of the
art a couple of years ago~\cite{Lutkenhaus2000}. The technique of
waveguided parametric
downconversion~\cite{Tanzilli2001,Banaszek2001,Jennewein2007,Kuklewicz2006}
and photon pair generation by four-wave mixing in dispersion-shifted
and photonic crystal fibres~\cite{Fulconis2007,Li2004} could provide
many improvements in the future.  By now, the brightness of PDC is
larger than required, cf.~\cite{U'Ren2004,Fulconis2005}.

\subsection{Detectors}
The detection process poses two requirements. Firstly, different
orthogonal signals need to be distinguished. Secondly, the detector
needs to announce if a signal arrived or not.

The decoding of orthogonally polarised signals depends on the
particular protocol, but can always be achieved by a polarising beam
splitter or an interferometer. Experimental imperfections can, for
instance, arise from a geometric misalignment of components. With a
certain probability, Bohr receives a logical ``zero'', but detects a
``one'' instead, and vice versa.  The error is conventionally referred
to as \emph{misalignment}, and the probability with which such an
error occurs is denoted by \(e_{\text{det}}\).  Note, that errors
which occur in the preparation stage can, as well as transmission
errors, be subsumed into detection misalignment.

The detection problem is usually more involved.  Although Planck might
have sent off a pulse, losses during the transmission can have led to
complete extinction, especially at longer distances. Owing to dark
counts, a detector can nevertheless click although no signal is
present.  The probability for this to happen is denoted by
\(p_{\text{dark}}\).  With current technology, it is hard to reliably
detect single quanta of light because of non-unit quantum detection
efficiency. A number of possibilities are available. Most commonly
avalanche photo diodes (APDs) are employed, but there are
alternatives, \eg, quantum dot detectors, visible light photocounters,
superconducting detectors etc.~\cite{Duvsek2006}. All suffer from
quantum efficiencies much below one.  In the following, we will denote
the probability to detect an incoming single photon with
\(\eta_{\text{det}}\). The typical value depends highly on the
wavelength used. While devices operating at \(800~\text{nm}\) provide
efficiencies of up to 80\%, the typical level for telecommunication
wavelengths (\(\approx 1550~\text{nm})\) is only around 5\%.

Click detectors with \(100\%\) quantum efficiency would allow to distinguish
between different orthogonal polarisations using a single detector: A click
represents one specific polarisation, and no click ensures that the orthogonal
polarisation was present. Under the influence of loss, this will not work
anymore: When a ``no click'' event takes place, it is impossible to
distinguish between the second polarisation mode and a lost signal.  Therefore
all practical QKD systems use two detectors to analyse the basis of the
transmitted signal.

\subsection{Information encoding}
Albeit arbitrary quantum systems can be used to encode information on them,
the use of light is a natural choice for present-day systems. Since binary
systems are the canonical fundamental of information science, the two-level
property is also carried into the quantum domain -- and called qbit
accordingly. Two orthogonal states are required to encode one classical bit
with values \(0\) and \(1\). The polarisation of a photon (which is equivalent
to a spin \(\frac{1}{2}\)-system) provides two orthogonal states, for instance
horizontal \(\ket{\leftright}\) and vertical \(\ket{\updown}\) polarisation.
It is possible to form arbitrary superpositions of these states. Physically,
the preparation of polarised light is a standard task easily achieved by
wave plates. On the detection side, a polarising beam splitter is
sufficient to distinguish between orthogonal polarisations.

Transmitting four differently polarised photons across fibres is 
challenging: Polarisation mode dispersion leads to polarisation
rotation. Although there are polarisation maintaining fibres which
preserve a single polarisation mode well, these cannot be used for QKD
because states with different, non-orthogonal polarisations need to be
transmitted. Fibre-based polarisation coding can in general only be
employed at short distances.

To remedy this drawback, an equivalent of a two-level system which is not
based on polarisation needs to be employed. Phase encoding~\cite{Gisin2002} is
mathematically equivalent to a two-level system, and thus to photon
polarisation.  By providing a phase reference to Bohr, Planck can prepare
signals with a defined phase, and Bohr can detect this phase using an
interferometric setup. Certain phase differences are assigned to one bit
value, and others to the conjugate value. One particular problem arising
here is phase mismatch of the interferometer, but this problem can be
controlled with current technology.

\subsection{Transmission}
There are basically two alternatives to transmit light from Planck to
Bohr: Over fibres, and across free space. It highly depends on the
wavelength which method is more favourable to deploy. For the
telecommunication wavelengths around 1550 nm, much research and
engineering effort has gone into optimising fibre
transmission. Current damping rates are at \(0.21~\text{dB/km}\). This
is quite close to the technological
limits~\cite{Gisin2002}\footnote{In principle, loss-less
  transmission would be possible by using teleportation. For this
  reason, we must assume that Einstein has quantum channels capable of
  transmitting signals without \emph{any} loss, since no fundamental
  physical reason prevents the existence of such channels.}, and huge
improvements are not to be expected.

The second wavelength region commonly employed is \(800~\text{nm}\), but
fibres in this region display large damping. Free space transmission is
therefore the most reasonable option. However, signal transmission only works
when a sight contact can be established, and weather conditions have a
significant impact on the attenuation. On a clear, calm night at best
weather conditions, the damping can be comparable to fibre transmission at
\(1550~\text{nm}\), but storm, clouds and mist can make damping rise to levels
of over \(20~\text{dB/km}\)~\cite{Duvsek2006}. It is crucial to observe that
the detection benefits at \(800~\text{nm}\) are compensated by the disadvantages
of transmission, and vice versa for \(1550~\text{nm}\).

\section{Quantum key distribution protocols}
A two-stage strategy is employed to establish a secret key between Planck and
Bohr which can be used to encrypt messages with perfect security.

\begin{figure}
  \centering\includegraphics[width=\textwidth]{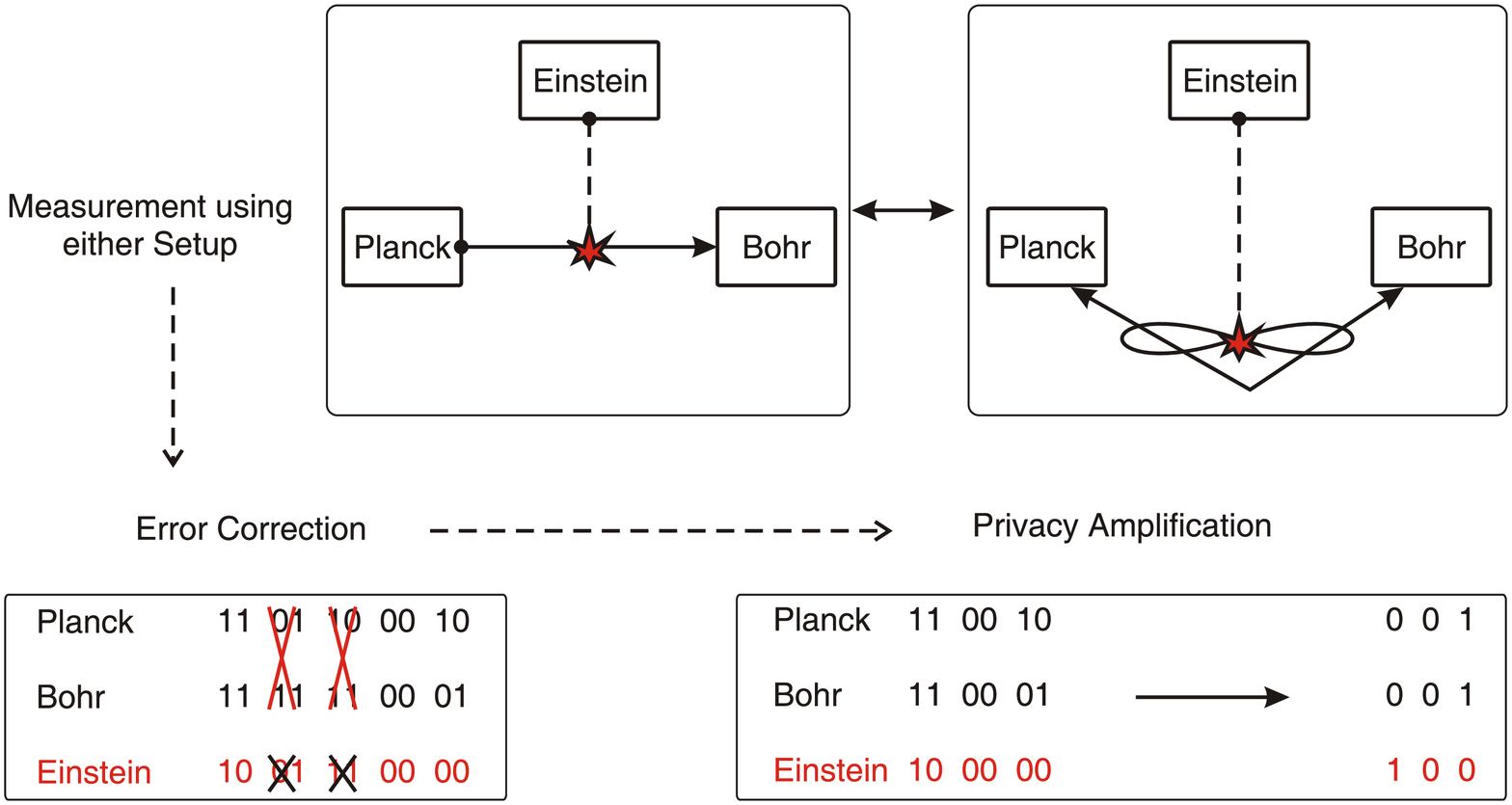}
  \caption{General structure of quantum key distribution. Planck
    provides a number of quantum mechanical signals to Bohr, and the
    eavesdropper Einstein tries to listen undetected. Employing
    various techniques described in this paper, Planck and Bohr can,
    however, generate a string of secret bits about which Einstein has
    no information whatsoever.  Starting from this key, classical
    mathematical methods can be employed to communicate in perfect
    security.}\label{fig:qkd_overview}
\end{figure}

\begin{itemize}
\item In the first stage, Planck transmits quantum mechanical signals to Bohr
  via a quantum channel. Since the signals obey the laws of quantum mechanics
  (especially the no-cloning theorem~\cite{Wootters1982}), measurements by
  Einstein will inevitably modify the state. These modifications manifest
  themselves as noise on the receiver side. If the noise exceeds a certain
  threshold, Einstein has potentially gained too much information about the
  signal, and the honest parties abort the protocol. Usually, the first stage
  of a QKD protocol is called \emph{quantum stage}.
\item The key shared between Planck and Bohr still contains errors, and
  Einstein may possess partial knowledge. Classical methods (error correction
  (EC) and privacy amplification (PA), see~\cite{Gisin2002,Duvsek2006}) are
  used in the second stage to remedy these. This stage can also include a
  sifting step, cf. Section~\ref{sec:BB84_intro}, in which inconclusive
  measurement results are removed.  Conventionally, the second stage is called
  \emph{classical stage} of a QKD protocol.
\end{itemize}

Once a perfect secret key between Planck and Bohr is established, a
classical method can be employed to perform the actual encryption: the
one-time pad, also called Vernam cipher. This method has been known
for a very long time~\cite{Vernam1926}, and provides perfect,
unbreakable security irregardless of any technological constraints.
The bits of a message \(m\) are added (naturally modulo 2) to the bits
of the secret key \(k\), generating the cipher text \(c = k \oplus
m\).  The encrypted message \(c\) is then sent to Bohr, and the plain
text can be recovered if and only if the key \(k\) is known: \(m = c
\oplus k\).

In this paper, we restrict our attention to discrete variable
protocols.\footnote{Continuous variable protocols where key distribution is
  achieved by employing conjugate quadratures for bright beams are
  an alternative. We refer the reader to the Review~\cite{Braunstein2005}
  for more information.} They use single quanta (or a very low number of
quanta) of the electromagnetic field which are most conveniently described in
a Fock space. Spin \(\frac{1}{2}\)-like properties are usually used to encode
and transmit information.  Obviously, this approach requires dark light beams
with very low intensities and provides corresponding experimental challenges,
but is well suited for security proofs. Discrete variable protocols come in
two flavours.

\begin{itemize}
\item In a \emph{prepare and measure} (p\&m) protocol, Planck
  prepares a quantum state and sends it to Bohr, who performs a measurement. A
  random choice of non-orthogonal encoding and measurement bases (see
  Section~\ref{sec:BB84_intro}) guarantees that Einstein cannot obtain perfect
  information about the transmitted states.
\item On the other hand, in an \emph{entanglement based} scheme, entangled
  photon pairs are utilised for key generation. One photon is given to Planck,
  and another to Bohr. Note, that the source of entangled states need not be
  located with the honest parties, and can even be placed under Einstein's
  control~\cite{Ma2007}!

  Planck and Bohr must ensure that they share a maximally entangled state.
  This can, for instance, be generated from a number of noisy non-maximally
  entangled states via entanglement purification and
  distillation~\cite{Gisin2002} (however, the physical operations required can
  effectively be replaced by EC and PA). Measurements on both sides will
  result in perfect correlations and Einstein cannot infer information
  at all.
\end{itemize}

It is interesting to observe that security proofs for p\&m schemes are often
founded on an entanglement based protocol operating on maximally entangled
states, followed by stepwise reduction to the p\&m situation,
cf.~\cite{Shor2000}. Experimental implementation differs in quite many details
though. We consider both, prepare and measure schemes as well as entanglement
based key distribution.

Classical communication between Planck and Bohr is necessary to perform various
tasks required for QKD. This communication is made publically available such
that Einstein cannot modify it once it has been sent off. Nevertheless, it remains
a problem to ensure that the origin of communication is really the intended
person.  This can be solved classically by using \emph{authenticated
  channels}. Since implementing an authenticated channel requires
a shared secret key, a causality dilemma arises: How can the key required for
authentication be transmitted?  A short initial secret key shared between
Planck and Bohr is required. Because of this dependency, QKD is 
thus referred to more exactly as \emph{quantum key growing}.

\subsection{BB84}\label{sec:BB84_intro}
BB84 is a p\&m scheme which relies on two sets of orthogonal bases, provided
by any two-level quantum system, for instance photon polarisation, phase
encoding, or even Spin \(\frac{1}{2}\)
systems~\cite{Duvsek2006,Gisin2002}. Set 1 consists of the two states
\(\shor\) and \(\svert\) with \(\braket{\updown}{\leftright} = 0\). Set 2
contains the states \(\ket{+} \equiv \sdiaga\) and \(\ket{-} \equiv \sdiagb\),
again with zero overlap. In every set, one state is identified with the
logical bit value zero, the other with the logical bit value one.

Since \(\ket{+}\) and \(\shor\) are non-orthogonal, their overlap is
non-vanishing: \(\braket{+}{\updown} = \frac{1}{\sqrt{2}}\). The same
non-orthogonality holds for any other combination of states from different
sets. Given a state from any of the two sets, the very laws of quantum
mechanics render it impossible to perfectly distinguish between the states!
Planck and Bohr proceed in the quantum stage as follows.

\begin{itemize}
\item Planck randomly chooses a bit value, and also randomly chooses one
  of the two sets to encode the bit. The encoded state is sent to Bohr
  \emph{without} announcing the set which is used for encoding. 
\item Bohr randomly chooses one of the two sets, and detects the
  signal in the appropriate basis.
\end{itemize}

After a number of bits have been transmitted this way, Planck and Bohr
announce which encoding/\hspace{0pt}measurement bases they have
chosen. Incompatible events where different sets were utilised are ignored,
and only compatible events are kept for further processing. This is commonly
referred to as \emph{sifting}. Assuming perfect hardware, Bohr's measured
results should completely agree with the states sent by Planck, and they can
easily check this by announcing a small number of sent/received values. If
Einstein has interacted with the signals, he will have introduced noise
because of the indistinguishability of non-orthogonal quantum states. This
leads to discrepancies of the bit values in the selected subset, and thus his
presence can be detected. We emphasise that this is impossible in a purely
classical scenario where signals can be measured without introducing
additional noise.

\subsection{SARG04}
Introduced two decades later than BB84, SARG04~\cite{Scarani2004}
keeps the tradition of naming QKD protocols by their inventors. The
same hardware components as for BB84 are required. The whole quantum
stage of SARG04 is identical to BB84. What differs, however, is how
bit values are extracted from the quantum measurements. The four
states available are now denoted by \(\ket{S_{0}}, \ldots,
\ket{S_{3}}\), and the relation \(\ket{S_{(n+m) \mod 4}} =
R^{n}\ket{S_{m}}\) (\(R\) denotes an apt rotation) holds. Four sets
\(\{ R^{k}\ket{S_{0}}, R^{k}\ket{S_{1}}\}, k \in \{0,1,2,3\}\) are
declared.

After the quantum stage is finished, Planck does not announce his
bases to Bohr, but only reveals one of the subsets that contains the
state which has been sent. If Bohr's measurement outcome is orthogonal
to one of the two states in the set, he can conclude that the other
state in the set has been sent, and he has obtained a conclusive
result. It is also possible to obtain non-conclusive results, namely
if the result is not orthogonal to all states in the set. After
repeating this procedure for all signals, Bohr announces where he has
found conclusive and inconclusive results, and the protocol commences
as for BB84.

\subsection{Decoy state variants}\label{sec:decoy}
The main advantage of QKD is the ability to detect eavesdroppers on
the line. If imperfect and lossy components (which cannot be avoided
in reality) are used to implement such systems, opportunities may
arise for Einstein which allow him to mimic the behaviour of
imperfections as a hideout (we will discuss usual weaknesses of QKD
systems and corresponding attacks further below).  It is thus
essential to test and characterise the utilised hardware not only when
a setup is installed, but also during the quantum stage of a running
protocol. One method to perform these tests is the decoy method. In
Section~\ref{sec:attacks}, we introduce an attack which can at present
only be detected by these means.

\subsubsection{Active decoy state selection}
Decoy states are an extension to p\&m QKD protocols. In addition to the
transmitted quantum states used for signal generation, Planck actively inserts
another set of states -- called \emph{decoy
  states}~\cite{Hwang2003,Lo2005,Li2006,Mauerer2007,Ma2005,Wang2007a,Wang2007,
  Wang2005b,Wang2005a} -- at random times. It is essential that the decoy
states have slightly different intensities than the signals, but are
indistinguishable from the proper signals in any other property. Since
Einstein cannot distinguish between signal and decoy states, he will perform
the same attacks on both, but owing to the slightly different intensities, the
attack has varying effects on signals and decoys.

Following the quantum stage, Planck announces the position of the
decoys in the signal stream, and Bohr announces his measurement
results. By comparing the error rates for signal and decoy states,
Planck and Bohr can infer photon-number resolved characteristics of
the channel without using photon-number sensitive hardware. This
yields very good estimates which are required for security proofs.

\subsubsection{Passive decoy state selection}\label{sec:tmd_decoy}
\begin{figure}[htb]
  \centering\includegraphics[width=\textwidth-1mm]{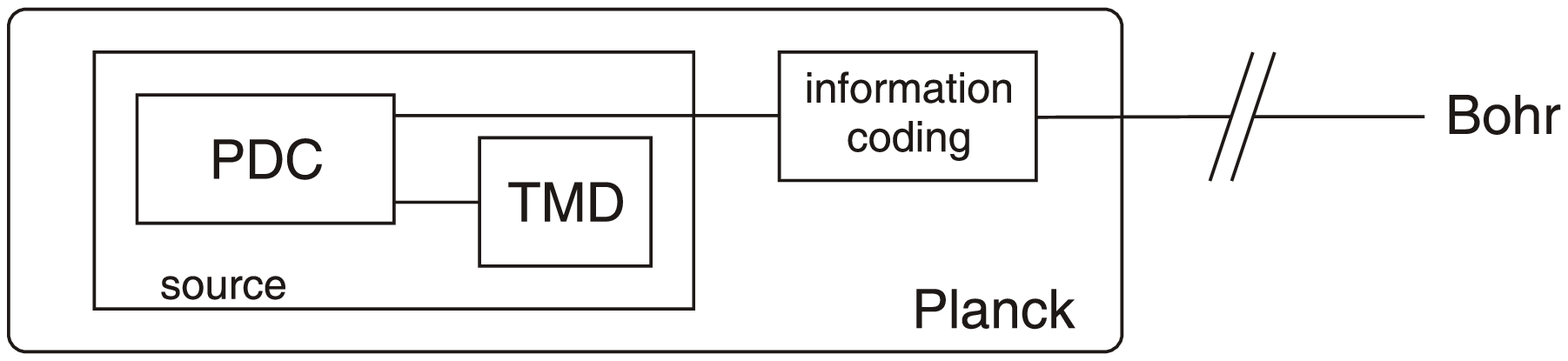}\\[3mm]
  \includegraphics[width=\textwidth-1mm]{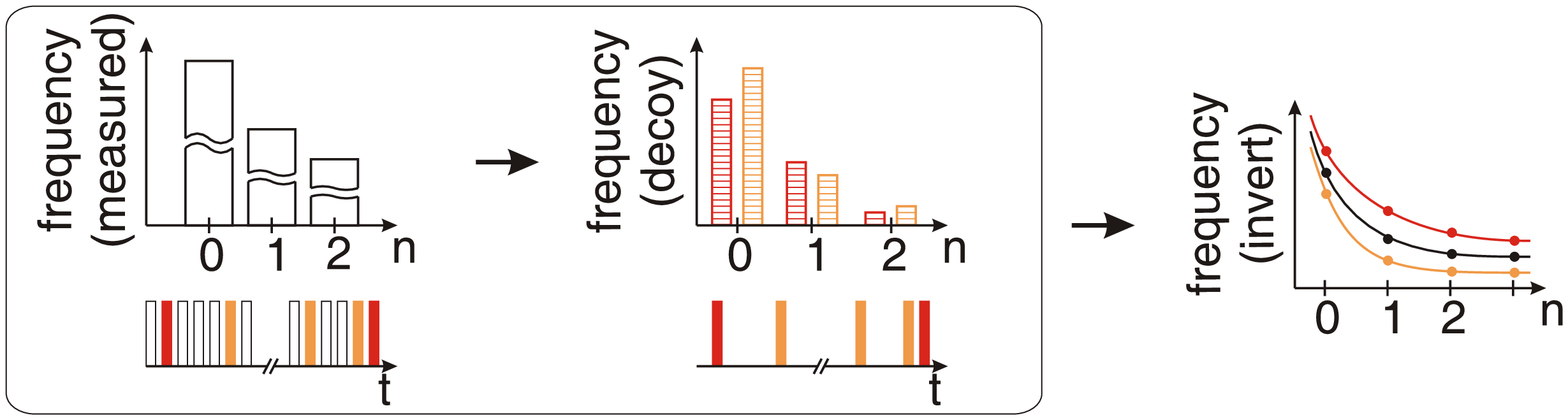}
\caption{Schematic implementation of the passive decoy QKD scheme (top), and
  how passive decoy state selection is performed (bottom). Refer to
  the text for further information.}\label{fig:tmd_decoy}
\end{figure}

It is crucial that signal and decoy states are experimentally
indistinguishable.  In a laser-based implementation, the exact
intensity can be difficult to control if it is varied for different
signals. Great care is required to ensure that intensity modifications
do not inadvertently modify other, unrelated parameters. However, it
is possible to perform the generation of decoy states in the classical
stage \emph{after} the quantum transmission is finished, thus
rendering a distinction between signal and decoy impossible \emph{in
  principle}. The setup required for this is depicted on the left hand
side of Fig.~\ref{fig:tmd_decoy}.

A PDC source on Planck's side generates two output states with strict photon
number correlations from an incident laser pulse. The time multiplexing (\ie,
approximately photon number resolving)~\cite{Fitch2003,Achilles2004} detector
(TMD) in one output arm is used to trigger on the spontaneous emission of PDC
photons and also to choose decoy states in the signal arm.  Information
encoding is performed on the signal arm, and afterwards, the state is sent to
Bohr via a quantum channel and analysed there as usual.

While photon number resolution cannot be obtained for single signals, it
becomes very reliable for a large number of measurements. An inversion process
can transform the measured statistics into the real, sent
statistics~\cite{Achilles2006}. This allows to select decoy states
\emph{after} the quantum stage~\cite{Mauerer2007}. Even Planck and Bohr do not
know which states are decoys and which states are signals during
transmission. Thus it is impossible \emph{in principle} for Einstein to
distinguish between signal and decoy states. The state selection is
demonstrated on Fig.~\ref{fig:tmd_decoy}.

The TMD records the measured photon number for each signal such that a
measured statistics can be obtained. The connection between sent statistics
\(\vec{p} = (p_{0}, p_{1}, \ldots, p_{N})\) and measured statistics
\(\vec{\varrho}\) is given by \(\vec{\varrho} =
\mathbf{C}\cdot\mathbf{L}\cdot\vec{p}\), where \(\mathbf{L}\) is a matrix
which compensates for lossy detection and \(\mathbf{C}\) accounts for
convolution effects in the detector. See Ref.~\cite{Achilles2004} for more
details.

This inversion process is not only possible for the whole set of transmitted
states, but also for (sufficiently large) subsets. To generate decoys, a
measured subset with a slightly different probability distribution as the
global set is selected. Inversion leads to a slightly different real
statistics than for the global set. The states in the selected subset
are then interpreted as decoys.

\subsection{Entanglement based schemes}\label{sec:entanglement}
PDC sources also allow to implement entanglement based QKD schemes.  While the
states generated by PDC sources are always photon number correlated, is is also
possible to include other degrees of entanglement, \eg, in frequency or in
polarisation~\cite{Kurtsiefer2001,Kwiat1995}. In Ref.~\cite{Ma2007}, a QKD
scheme based on entangled PDC states was introduced.  The state emitted from
the source has the form
\begin{equation}
  \ket{\psi} \propto
  \sum_{n=0}^{\infty}p(n)\sum_{m=0}^{n}(-1)^{m}\ket{(n-m)\updown,m\leftright}\ket{m\updown,(n-m)\leftright}. 
\end{equation}
For instance, \(n=1\) delivers the Bell state
\(\ket{\psi_{1}}\propto\ket{\updown}\ket{\leftright} -
\ket{\leftright}\ket{\updown}\).  A crucial point is that the source
needs not be placed on Planck's side, but can also reside in the
middle between Planck and Bohr. While this may sound insecure at a
first glance, it is indeed possible to show that this method is
completely secure if Planck and Bohr can proof the entanglement of the
state~\cite{Curty2004}.

Although entanglement based systems tend to provide a better theoretical
performance than p\&m implementations, their physical realisation can be
challenging, cf.~p.~176 in~\cite{Gisin2002}.

\section{Security proofs}
The basics of QKD protocols are easy to understand, but proving
unconditional security has shown to be very hard when imperfections are
included. The first proof by Mayers\cite{Mayers1998}, for instance, required
56 pages filled with calculations! Nowadays, proofs which employ rather
different techniques are available. We should like to emphasise two particular
methods.

\begin{itemize}
\item Information-theoretic techniques (ITT), most notably developed
  in~\cite{Renner2005}, provide mathematically sophisticated methods to prove
  the security of QKD under very general assumptions.  However, involved
  methods~\cite{Kraus2005} are required to compute actual numerical values for
  key rates and alike.
\item The security proof by Gottesman, Lo, L{\"u}tkenhaus, and Preskill
  (GLLP,~\cite{Gottesman2004}) might be less elegant than the aforementioned
  method, but facilitates easy calculation of actual performance data.
\end{itemize}

Results and performance achieved by GLLP-type methods are approximately
identical with ITT results, and have a more direct and accessible physical
interpretation. We will therefore concentrate on these in the following.
Note, that all proofs rely on the law of large numbers. This requires
transmission of a nearly infinite number of pulses. For obvious reasons, this
is hard to achieve in practice. The key rates for comparably small numbers of
transmitted signals (\(\approx 10^{4}\)) will thus decrease in comparison to
the numbers presented here, cf.~\cite{Meyer2006} for a detailed analysis.

\subsection{Possible attacks}\label{sec:attacks}
Owing to the historical development of security proofs, three different types
of attacks are conventionally distinguished.

\begin{itemize}
\item \emph{Individual attacks} restrict Einstein to attaching an independent
  probe system to one signal after another, and measuring these probes
  also one after another.
\item \emph{Collective attacks} similarly restrict Einstein to
  attaching an independent probe to each pulse, but allow him to
  measure several probes collectively.
\item \emph{Coherent attacks} are the most general form where Einstein can process
  an arbitrary number of qbits at one time.
\end{itemize}

We mention that all rate formulae used in the remainder of this paper hold
against coherent attacks.

One particular example for an individual attack often employed in proofs is an
intercept-resend attack. Einstein measures each signal coming from Planck. He
stores the result, and prepares a fresh pulse with his measurement result
which is sent to Bohr. Assuming the BB84 protocol, only 50\% of his bases
choices will be compatible with the signal. These measurements lead to the
correct result. The remaining 50\% of the results will be random. Bohr finds
that 25\% of all signals are erroneous.  This is no problem if the hardware
induces an error rate below 25\%.  The protocol is aborted if this threshold
is exceeded.  If the errors caused by the hardware alone are above
25\% percent, Planck and Bohr will not be able to detect an intercept-resend
attack, and security cannot be guaranteed any more. The upper bound
on the distance shown in Fig.~\ref{fig:dist_example} actually arises due to
the intercept-resend attack. It should however be noted that this is not the
smallest upper bound, cf., for instance, Ref.~\cite{Moroder2006,Moroder2006a}
on how to compute smaller bounds.

One particularly powerful attack in p\&m schemes is photon number
splitting~\cite{Lutkenhaus2002}. After replacing the lossy quantum
channel by a loss-less one, Einstein performs a non-destructive
measurement of the photon number and then adapts her action
accordingly. He blocks all single photon pulses. For multi-photon
pulses, he strips off one photon, stores it, and sends the remaining
ones to Bohr. After Planck has announced his encoding bases, Einstein
measures the stored photons and obtains full information. This
strategy does not increase the error rate at all if the photon number
distribution sent by Einstein is identical to the distribution which
would have been expected after transmission through a lossy
channel. In fact, the attack can only be detected in p\&m schemes if
the decoy method is employed.

\subsection{Anatomy of GLLP-style proofs}\label{sec:gllp_proofs}
We now discuss the results of the GLLP security proof in order to 
apply it to practical QKD setups to estimate the performance of
such systems.  Without going into
details of the derivation (which can be found in~\cite{Gottesman2004}), we
analyse the contributions to the rate equation. An adaption of the GLLP lower
bound on the secret key rate to BB84 with decoy states and one-way
communication as derived by Lo and coworkers~\cite{Lo2005} in 2005 is given by
\begin{equation}
  S_{\text{one-way}} \geq \underbrace{qQ_{\bar{N}}}_{\text{\hbox to0pt{\hss
        Conclusive signals \hss}}}
  \big(\underbrace{-f(E_{\bar{N}})H_{2}(E_{\bar{N}})}_{\text{Error correction}} + 
  \underbrace{\Omega\cdot
    \left(1-H_{2}(e_{1})\right)}_{\text{Privacy 
      amplification}}\big).\label{eqn:gllp_1locc}
\end{equation}
Note that the value of \(S\) denotes the key generation rate, \ie, the
fraction of secret bits which can be extracted per signal. The bit rate per
second can be obtained by multiplying \(S\) with the repetition rate of the
scheme. \(\bar{N}\) specifies the mean photon number of the source in use.
\(E_{\bar{N}}\) is the overall bit error rate, and \(e_{n}\) the photon number
resolved bit error rate.  We emphasise that the formula holds for sources with
arbitrary photon number distributions.

First of all, secret bits can only be extracted from events where Bohr has
obtained a conclusive result. Cases where signals are lost in the channel are
excluded by the prefactor \(Q_{\bar{N}}\). This factor denotes the fraction
between successful detection events on Bohr's side and the number of pulses
sent by Planck. The rate of conclusive single-photon events over all
conclusive events is given by \(\Omega = \frac{Q_{1}}{Q_{\bar{N}}}\).  Since
the detection bases need to be identical to obtain a conclusive result, the
sifting factor \(q\) specifies the probability that Planck and Bohr choose
identical bases.  The sifting stage is not necessary, though, if a very large
number of signals are transmitted, cf.~\cite{Lo2005c}. This also holds for any
other p\&m protocol considered in the following. Two operations need to be
performed on the raw bits (a visual summary of these operations can be found
in Fig.~\ref{fig:qkd_overview}).

\begin{itemize}
\item Error correction ensures that Planck and Bohr share a completely
  correlated string of zeroes and ones. This can be achieved by classical
  error correction algorithms (cf.~Ref.~\cite{Gisin2002,Duvsek2006}). A certain
  number of bits needs to be sacrificed to perform the correction. The
  Shannon entropy \(H_{2}(E_{\bar{N}})\) quantifies how many. Since the
  Shannon limit is not constructive, practical codes will perform less
  efficient. This is accounted for by a compensation factor \(f(E_{\bar{N}}),
  f\geq 1\).
\item Although Planck and Bohr share an identical key after error correction,
  Einstein may still be correlated with the key. To remove these correlations,
  privacy amplification~\cite{Gisin2002,Duvsek2006} is employed. Since only
  single photon signals are secure, \(\Omega\) is the base quantity for
  privacy amplification.  Additionally, only error-free single photons can
  contribute, which is accounted for by the factor \(1-H_{2}(e_{1})\).
\end{itemize}

While error correction can be performed equally efficient irregardless of one-
or two-way communication, PA is more powerful if Planck and Bohr can perform
bidirectional classical communication.  The rate equation which includes
two-way communication is as follows~\cite{Gottesman2002,Ma2006a}.
\begin{equation}
  S_{\text{two-way}} \geq q\tilde{r}_{\text{B}}Q_{\bar{N}}
  \left(-f(\tilde{E}_{\bar{N}})H_{2}(\tilde{E}_{\bar{N}}) + 
    \tilde{\Omega}\cdot\left(1-H_{2}(\tilde{e}_{1,p})\right)\right).
  \label{eqn:gllp_2locc}
\end{equation}
Eq.~\ref{eqn:gllp_2locc} is structurally similar to Eq.~\ref{eqn:gllp_1locc},
but differs in some details. First of all, bidirectional PA works in steps on
the classical bit string.  In each step, roughly half of the total bits
survive, but Einstein has less information on the shortened bit string. After
\(n\) PA steps, \(r_{B} \approx 2^{-n}\). Additionally, symmetry between bit
(\(e_{n,b}\)) and phase (\(e_{n,p}\)) error rates is lost.  Bidirectional PA
reduces the overall error rates \(E_{\bar{N}}\) as well as \(\Omega\), and
makes the combination \(e_{n,b}\) and \(e_{n,p}\) more favourable. Quantities
with a tilde represent the new values after performing the PA steps. We do not
want to list the transformation equations, but refer the reader to
Ref.~\cite{Ma2006a}.

\subsection{Koashi-Preskill proof for entangled states}
The key generation rate formula for the entanglement-based PDC QKD as
introduced in Section~\ref{sec:entanglement} is not based on the GLLP
proof. Instead, Ma and co-authors~\cite{Ma2007} founded their proof on a
technique introduced by Koashi and Preskill~\cite{Koashi2003}. The resulting
formula is as follows
\begin{equation}
  S \geq q\left(Q_{\bar{N}}[1-f(E_{\bar{N}})H_{2}(E_{\bar{N}})-H_{2}(E_{\bar{N}})]\right).
  \label{eqn:ent}
\end{equation}
Again, the structure is similar to Eq.~(\ref{eqn:gllp_1locc})
and~(\ref{eqn:gllp_2locc}), with one important difference. The error rates need
not be known for single photons, it suffices to know the overall error
rate. Intuitively, one reason for this is that entanglement-based schemes are
not affected by photon-number splitting operations, cf.~Ref.~\cite{Gisin2002}.


\section{Performance analysis}
Having introduced various protocols and how experimental imperfections are
included in their security proofs, we now discuss how these imperfections
influence the performance quantitatively.  We will compare the benefits which
arise when the individual components of a QKD system are improved. This will
aid experimental researchers in finding where physical and technical
improvements will pay off most.

\subsection{Numerical methods}
To facilitate systematic comparison of various QKD schemes, we have developed
a universal simulation tool which performs the necessary numerical
computations. Care was taken to ensure that no special knowledge about
security proofs or the numerics associated with them is necessary.  A
graphical user interface enables to conveniently obtain all data of
interest. The tool will be made available for download on our group's web
page\footnote{\texttt{http://www.optik.uni-erlangen.de} \(\rightarrow\) Max
  Planck Junior Group IQO \(\rightarrow\) Quantum Cryptography.}.

\subsection{Comparison between protocols}
\begin{figure}
  \centering\includegraphics[width=0.75\textwidth]{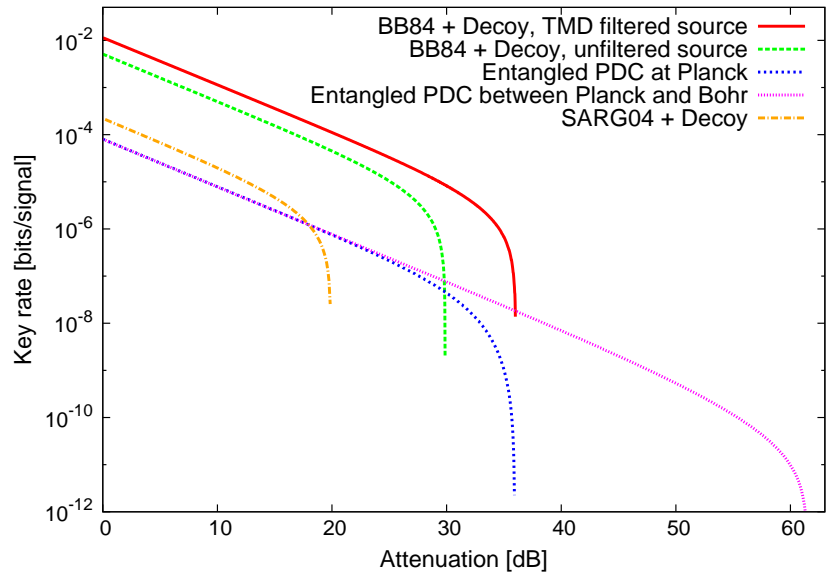}
  \caption{Comparison between various QKD protocols. The plot depicts lower
    bounds on the secret key rates.}\label{fig:protocols}
\end{figure}

First of all, it is desirable to know how the protocols perform compared to
each other when the same hardware is employed. Fig.~\ref{fig:protocols}
provides a series of performance plots for all protocols discussed in this
paper. All use the hardware described in the GYS experiment~\cite{Gobby2004}.

The largest distance is achieved by the entangled PDC protocol, which exceeds
the tolerable loss of other protocols by more than \(25~\text{dB}\). The
achievable key rate, however, is two orders of magnitude lower than for BB84
decoy protocols with and without source filtering. It can also be argued that
entanglement based protocols are harder to implement than p\&m protocols,
cf.~\cite{Gisin2002}.

Using SARG04 instead of BB84 does not provide any advantages when the security
proofs utilised in this paper are employed.\footnote{Note, however that this
  can change when different security analyses are employed. With the methods
  introduced in Ref.~\cite{Scarani2004}, SARG performs better than BB84, but
  the analysis does not employ decoy states.} Both, secret key rate and secure
distance are smaller than for BB84 with decoy states. For these reasons, we
focus on the BB84 decoy protocol in the following.

\subsection{Quantitative influence of imperfections}
\begin{figure}
  \centering\includegraphics[width=\textwidth]{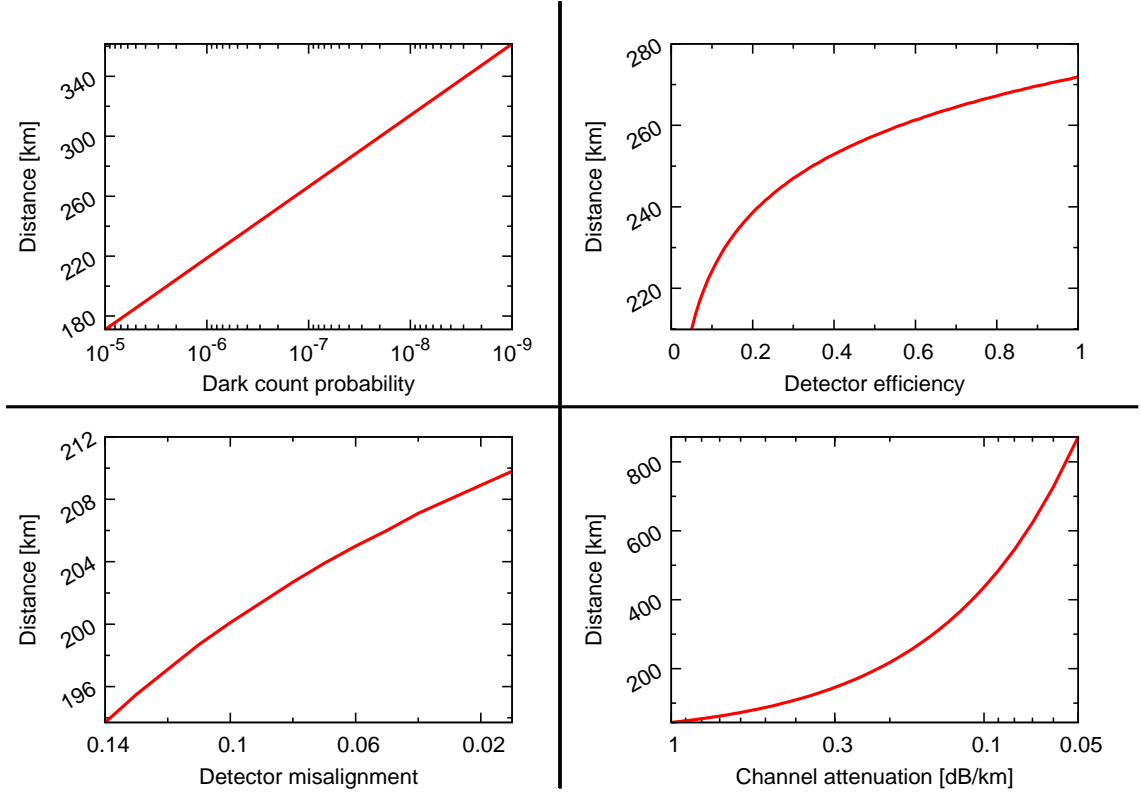}
  \caption{Comparison of achievable secure distances when experimental
    imperfections are varied. The standard parameters are \(p_{\text{dark}}\)
    = \(1.7\cdot 10^{-6}\), \(\alpha\) = \(0.21~\text{dB}/\text{km}\),
    \(e_{\text{det}}\) = \(3.3\cdot 10^{-2}\), \(q\) = \(1\), \(\lambda\) =
    \(1550~\text{nm}\), \(\eta_{\text{det}}\) = \(4.5\cdot 10^{-2}\), and the
    protocol is decoy based BB84. Refer to the text for further
    information.}\label{fig:smallplots}
\end{figure}

Before we discuss any specific scenarios, we would like to present a general
summary how the achievable secure distance changes with experimental
imperfections. The corresponding graphs can be found in
Fig.~\ref{fig:smallplots}.

Little improvement can be expected when the detector misalignment is
improved. The current experimental standard is \(e_{\text{det}} = 0.03\). If
this value were reduced to \(0.02\), the increase in secure distance is only
around \(3~\text{km}\). Even a very large misalignment of \(10 \%\) causes
only approximately \(8~\text{km}\) decrease.\footnote{Note, that for simplicity
  we neglect weakly basis dependent attacks~\cite{Gottesman2004} here.} This
also implies that slightly different efficiencies or dark count rates of the
two detectors can be mostly neglected.

Improving channel attenuation would be highly beneficial as the bottom
right graph in Fig.~\ref{fig:smallplots} demonstrates. However, the
damping for telecommunication fibres is quite close to the optimal
technological possibilities as we have explained before. Bad weather
drastically increases free space damping and thus shortens the secure
distance considerably. Furthermore, the constantly varying conditions
need to be monitored during signal transmission.

Moderate possibilities for improvement are provided by optimised detection
efficiency. Observe the top right graph in
Fig.~\ref{fig:smallplots}. Improving \(\eta_{\text{det}}\) from, say, \(10
\%\) to \(60 \%\) results in a increased secure distance of approximately 
\(30\ \text{km}\).

We find that the best possibility to increase the secure distance is to
suppress dark counts, as can be seen in the top left side of
Fig.~\ref{fig:smallplots}.  Decreasing this rate by one order of magnitude
delivers \(40\) additional secure kilometres. Considering the development over
the last years, it is realistic to assume that better time gating and new
detectors with increased single photon sensitivity will allow to improve
\(p_{\text{dark}}\) by approximately three orders of magnitude.

\section{Example scenarios}
Let us finally turn our attention to four practical scenarios. We base our
analysis on a number of QKD experiments performed in recent years.  The
parameters of components employed for them is collected in
Table~\ref{tab:qkd_params}.

\newcommand{\myhline}{\hline}
\renewcommand{\tabcolsep}{1.6mm}
\begin{table}
\begin{small}
    \hbox{\begin{tabular}{llllllllll}
      Ref. & Protocol & \(\lambda\) [nm] & Mean photon \# & \(\alpha\) & \(\eta_{\text{det}}\) & 
      \(e_{\text{det}}\) & \(p_{\text{dark}}\) & Enc. & QBER \\\myhline
      \cite{Dynes2007} & BB84 + & $1550$ & signal: $0.55$        & unspec. &
      $5.62$\% & unspec. & $1.4\cdot 10^{-4}$ & Pol. & unspec.\\
      & Decoy  &        & decoy: vac., $0.098$  &         &
      &         &                    &      &        \\\myhline
      \cite{Gobby2004} & BB84 & $1550$ & $0.0042$ -- $0.046$ & $0.21$ &
      $4.5$\% & $3$\% & $8\cdot 10^{-7}$ & Phase & $3$ -- $6$\%\\
      &      &        &                    & dB/km  &
      &       &                  &       &            \\\myhline
      \cite{Peng2007} & BB84 + & $1550$ & signal: $0.6$       & $0.2$ &
      unspec. & unspec. & $6.7\cdot 10^{-6}$ -- & Pol. & $3.2$\% --\\
      & Decoy  &        & decoy: vac., $0.2$  & dB/km &
      &         & $9.2\cdot 10^{-6}$   &      & $3.6$\%\\\myhline
       \cite{Yin2007} & BB84 + & $1550$ & signal: $0.6$ & unspec. &
      unspec. & unspec. & $5\cdot 10^{-7}$ & Phase & $1$ -- $2$\%\\
      & Decoy  &        & decoy: $0.2$  &         &
      &         &                  &       &            \\\myhline
      \cite{Yuan2007} & BB84 + & $1550$ & signal: $0.425$ & $4.7$~dB/ &
      $5.6$\% & unspec. & $9.4\cdot 10^{-5}$ & Pol. & $1.72$\%\\
      & Decoy  &        & decoy: $0.204$  & $25.3$~km &
      &         &                    &      &         \\\myhline
      \cite{Resch2005} & Entangled & $810$ & unspec. & $1.4$\%/ &
      $15$\% & unspec. & $800~\text{s}^{-1}$ & Pol. & $9.9$\%\\
      &         &       &         & $7.8$~km  &
      &         &                    &      &        \\\myhline
      \cite{Schmitt-Manderbach2007} & BB84 + & $850$ & signal: $0.27$      & $24$~dB/ &
      $10$\% & $3$\% & unspec. & Pol. & $6.48$\%\\
      & Decoy  &       & decoy: vac., $0.39$ & $144$~km  &
      &       &         &      &        \\\myhline
 \end{tabular}\hss}
\end{small}
  \caption{Summary of parameters used in recent QKD experiments.}
  \label{tab:qkd_params}
\end{table}

\subsection{Radio link free space transmission}
\begin{figure}
  \centering\includegraphics[width=\plotwidth]{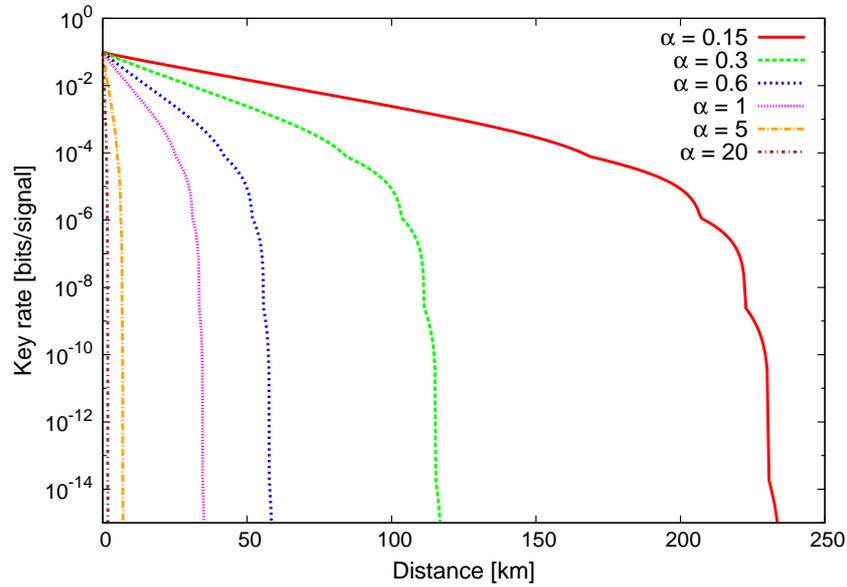}
  \caption{Performance analysis for radio link based QKD systems. The plot
    depicts lower bounds on the secret key rates. Refer to the text for more
    information. The connection between attenuation per length \(\alpha\),
    channel length \(L\), and loss \(\eta\) is given by \(\eta
    =10^{-\alpha L/10}\).}\label{fig:funk}
\end{figure}
Short distance QKD between buildings in a city, for instance, can conveniently
be realised via radio links operating at \(800~\text{nm}\). Because of stray
light at daytime, such schemes will only be operational at night. Another
crucial factor is signal attenuation. Weather conditions impose serious
differences in loss per length. This in turn has a considerable effect on the
achievable key generation rate. Fig.~\ref{fig:funk} shows the performance
for a wide range of attenuation values.  While the key rate at best conditions
gives values in the satisfactory range of \(10^{-3}\) bits per pulse,
performance will rapidly decrease with increasing attenuation. In bad weather,
the channel loss can be in the range of \(20~\frac{\text{dB}}{\text{km}}\),
and the secure distance will only be around \(10~\text{km}\), which is often
not sufficient for any communication at all.

\subsection{Satellite-based QKD}
\begin{figure}
  \centering\includegraphics[width=\plotwidth]{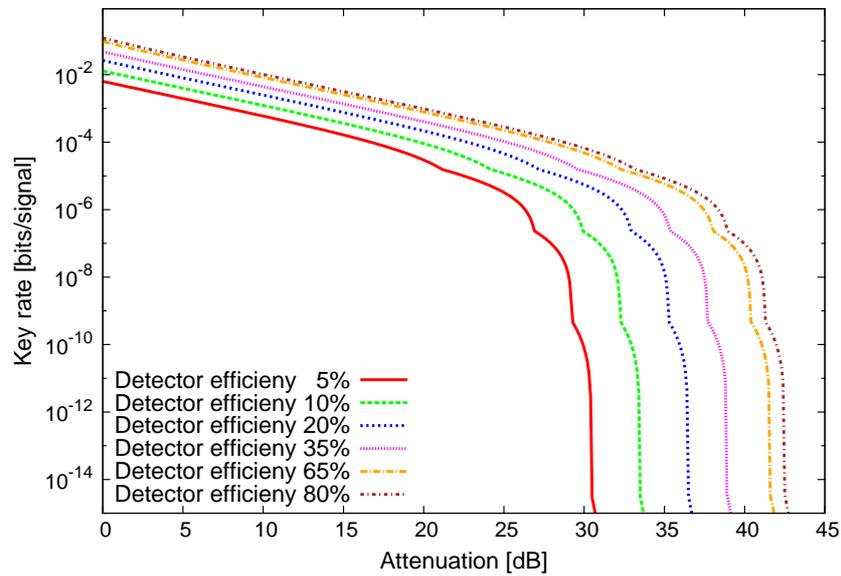}
  \caption{Influence of detector efficiency for a satellite link on the secure
    key generation rate.}\label{fig:satellite}
\end{figure}
Recently, experimental free space QKD
implementations~\cite{Schmitt-Manderbach2007} have reached transmission
distances which would allow to communicate with geostationary
satellites.\footnote{The attenuation between earth and a geostationary
  satellite is around \(35~\text{dB}\).} This is an important break-through
because it would enable a world-spanning secure QKD network. Since
transmission losses from the earth into space cannot be actively influenced,
the detector efficiency is the main parameter which can be improved.  The
implementation from op.~cit.~employed a detector with a quantum efficiency of
\(10 \%\). Fig.~\ref{fig:satellite} shows how the secure key rate
would develop with better detection efficiency (clearly, the absolute value of
the maximal transmission distance is not of interest because the path length
from a ground station to the satellite is fixed). While the quoted detection
efficiency barely suffices to reach the desired distance, a increase by \(20\)
to \(30\) absolute percent would move the rapidly declining part of the key
rate to distances beyond the satellite. The desired transmission distance now
lies in the slowly falling part, resulting in a secret key rate increased by
many orders of magnitude.

\subsection{Long-Distance fibre based transmission}
\begin{figure}
  \centering\includegraphics[width=\plotwidth]{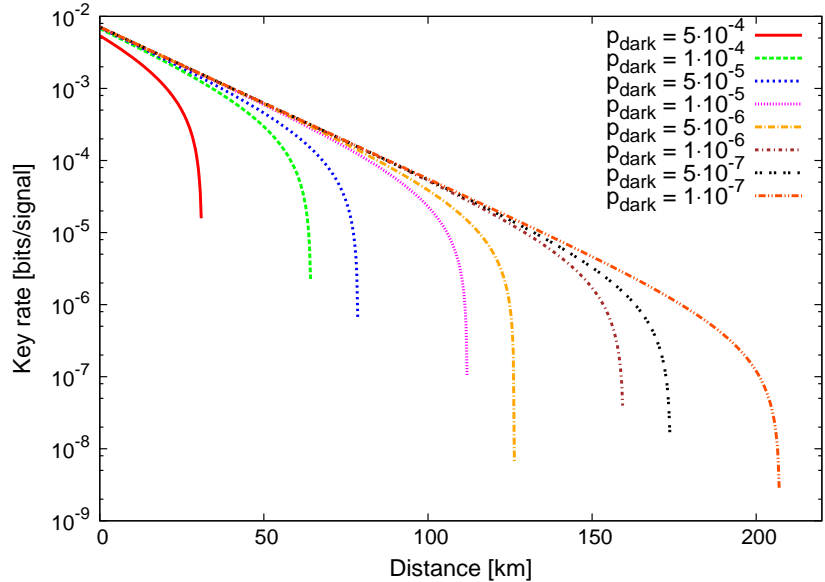}
  \caption{Influence of detector efficiency in a fibre based system at
    \(1550~\text{nm}\) on the secure key generation rate.}\label{fig:fibre}
\end{figure}

The most promising candidate for intermediate to long range
terrestrial communication are fibre-based systems operating at \(1550
\text{nm}\).  Several experiments in this regime have been performed
recently~\cite{Yuan2007, Yin2007,Peng2007, Dynes2007,Gobby2004}.  The
key parameter which provides room for improvement is the dark count
probability. While the detectors in the quoted papers provide
probabilities in the range between \(10^{-5}\) and \(10^{-6}\), it can
reasonably be expected that the probability will be as low as
\(10^{-9}\) in future detectors. Fig.~\ref{fig:fibre}
demonstrates that while the rates are unaffected by the dark count
rates, the secure distance increases considerably with decreasing dark
count probability.

\subsection{Passive decoy QKD with two-frequency downconversion}
\begin{figure}
  \centering\includegraphics[width=\plotwidth]{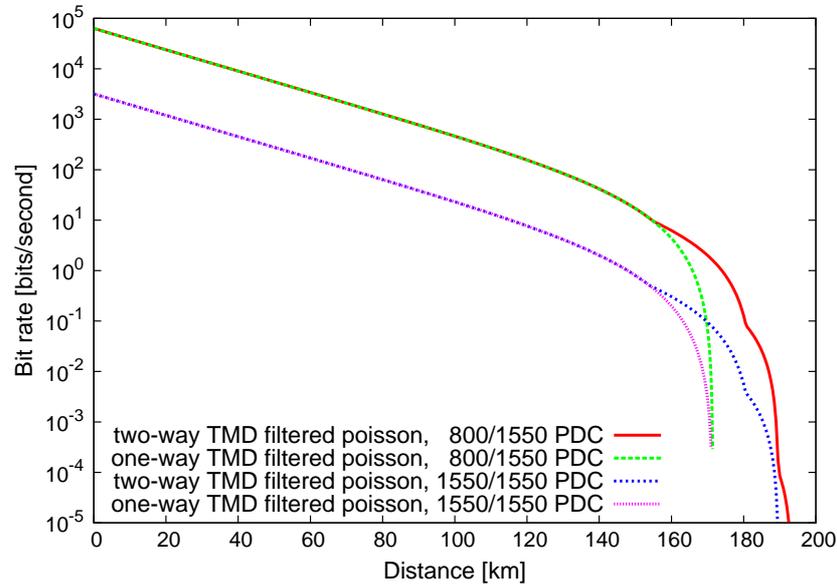}
  \caption{Performance prediction for a mixed-wavelength passive decoy
    scheme. The signals emitted at \(1550~\text{nm}\) are transferred over a
    quantum channel, while the idler photons at \(800~\text{nm}\) are used for
    triggering and passive decoy selection with good quantum efficiency. The
    secure key rate in comparison to a passive decoy scheme with signal and
    idler wavelengths at \(1550~\text{nm}\) is considerably
    increased.}\label{fig:new_scheme}
\end{figure}

Transmission over fibres is advantageous for wavelengths in the \(1550
\text{nm}\) regime, but good detectors are only available for \(800
\text{nm}\). Particular advantages gained by employing one wavelength region
are always negated to some extent in conventional QKD setups. A proposed setup
is a variation of the scheme introduced in Section~\ref{sec:tmd_decoy}. It
tries to overcome this limitation by picking advantageous features from both
worlds. As long as energy and momentum are conserved via
(quasi)-phase-matching, downconversion sources can well be built to emit
signal and idler photons at different wavelengths. Signals are emitted at
\(1550~\text{nm}\) and transmitted over a quantum channel with little
loss. The idler arm emits photons at \(800~\text{nm}\); these can be detected
by the TMD with good quantum efficiency (around \(70 \%\)).
Fig.~\ref{fig:new_scheme} presents the performance prediction for the proposed
scheme.  The graph is based on a source repetition rate of \(20~\text{MHz}\),
Ref.~\cite{Coldenstrodt-Ronge2007} has shown that this is the maximum for TMD
detection with current technology.  While the maximal secure distance is
identical in both cases, the mixed wavelength scheme provides a bit
transmission rate which is one order of magnitude higher than for identical
wavelengths.

\section{Conclusions}
In summary, we have presented a selection of discrete variable quantum key
distribution systems. After analysing the structure of the rate formulae
obtained from security proofs, we have given a systematic comparison of
experimental tunables and their influence on the performance. The areas where
experimental improvements will bring maximal benefit have been
exposed. Finally, we have performed an analysis of four practical
scenarios under consideration of current experimental possibilities.

\begin{acknowledgement}
  We would like to thank Benjamin Brecht for his most valuable and
  long-standing help with image preparation.
\end{acknowledgement}

\def\bstname{adp}
\bibliographystyle{adp}
\bibliography{qkd}
\end{document}